\documentclass[
aps,%
12pt,%
final,%
notitlepage,%
oneside,%
onecolumn,%
nobibnotes,%
nofootinbib,
superscriptaddress,%
noshowpacs,%
centertags]%
{revtex4-1}
\RequirePackage{graphicx}

\def\npb#1#2#3{Nucl. Phys. B #1 (#3) #2}

\def\prl#1#2#3{Phys. Rev. Lett. #1 (#3) #2}

\newcommand{\order}[1]{ \mathcal{O} \left( #1 \right) }

  \newcommand{\lqcd}{\Lambda_{QCD}}

\def\refitem#1{\relax}

\begin{document}
\title{How large is ``large $N_c$'' for Nuclear matter?}

\author{\firstname{G.} \surname{Torrieri}}
\email{torrieri@th.physik.uni-frankfurt.de}
\affiliation{FIAS, JW Goethe Universitat, Frankfurt, Germany}

\author{\firstname{I.} \surname{Mishustin}}
\affiliation{FIAS, JW Goethe Universitat, Frankfurt, Germany}
\affiliation{Kurchatov Institute, Russian Research
Center, Moscow
123182, Russia.}

\begin{abstract}
We argue that a so far neglected dimensionless scale, the number of neighbors in a closely packed system, is relevant for the convergence of the large $N_c$ expansion at high chemical potential.   It is only when the number of colors is large w.r.t. this new scale ($\sim \order{10}$) that a convergent large $N_c$ limit is reached.  This provides an explanation as to why the large $N_c$ expansion, qualitatively successful in in vacuum QCD, fails to describe high baryo-chemical potential systems, such as nuclear matter.  It also means that phenomenological claims about high density matter based on large $N_c$ extrapolations should be treated with caution.  This work is based on \cite{largenc}
\end{abstract}

\maketitle

\section{Introduction: QCD matter at large $N_c$}
Strongly interacting matter at moderate ($\sim$ the confinement scale) quark chemical potential $\mu_q$ and moderate temperature $T$ has recently received a considerable amount of both theoretical and experimental interest.   Such matter can hopefully be produced in heavy ion collisions \cite{low1,low2,low3,low4}, and is thought to exhibit a rich phenomenology: critical points \cite{critical},  instabilities \cite{spino}, precursors to color superconductivity \cite{blas}, separation between chiral symmetry and confinement \cite{ratti,dirkq,sasaki1} chirally inhomogeneous phases \cite{chiralspiral1,buballa1}, new phases \cite{quarkyonic} etc.

These conjectures are, however, extraordinarily difficult to quantitatively explore in a rigorous manner.
The quark chemical potential $\mu_q$ is nowhere near the asymptotic freedom limit where perturbative QCD can be used \cite{jansbook}.   It is, however, way too high for existing lattice-based approaches, dependent on $\mu_q/T \ll 1$, to work \cite{latticebook}.

Perhaps the only relevant quantity with can be uncontroversially be called ``a small parameter'' (albeit not so small in the real world!) is $1/N_c$, where $N_c$ is the number of colors \cite{thooft,witten}.   While the large asymptotically $N_c$ theory shares with QCD asymptotic freedom for hard processes and confinement for soft ones (separated by an energy scale $\lqcd \sim 250$ MeV \cite{pdg}, independent of $N_c$) , the  $N_c$ scaling of different observables can be used to establish a model-independent hierarchy.  Thus, the shape of the phase diagram can be said with relative certainty to look like Fig. \ref{phasediag}:   Phases I and III are, respectively, the familiar confined chirally broken Hadron gas (where pressure $\sim N_c^0$) and the deconfined chirally-restored quark-gluon plasma (where pressure $\sim N_c^2$). Since at large $N_c$ gluon loops dominate over quark loops,  the critical  temperature  $ \sim N_c^0 \lqcd$, and the critical chemical potential necessary for deconfinement is very high,
$\mu_q \sim N_c^2 \lqcd$.  

Consequently, in the large $N_c$ limit, the phase transition line becomes horizontal for moderate $\mu_q$.  In this limit the transition between zero baryonic density and finite baryonic density matter is infinitely sharp at $N_c \mu_q \sim m_B \sim N_c \lqcd$ \cite{quarkyonic}, since the baryon density $\sim \exp\left[ -N_c\left( \lqcd - \mu_q\right) \right]$ goes to zero exponentially with $N_c$ for chemical potentials less than the baryonic mass.   Thus, a new phase (II) emerges where the nuclear density is $\order{1} \lqcd^3$, parametrically much less  then that required for deconfinement, $\order{N_c} \lqcd^3$, but much more then the density of vacuum QCD $\order{\exp\left[-N_c\right]} \lqcd^3$.   

Naively, since $\mu_q \sim \lqcd$ is nowhere near the chemical potential required for deconfinement, this phase should just be that of dense nuclear liquid (the large $N_c$ limit of the nuclear liquid, well-studied theoretically and experimentally \cite{liq0,liq1,liq3,liq5}), where nucleons are close to touching each other, yet confinement is still there and degrees of freedom are baryons and mesons.   When this phase is considered at variable $N_c$, it is naively expected that the energy density $\sim N_c$ (since the mass of each baryon $\sim N_c$), but pressure and entropy density $\sim N_c^0$, since the energy is locked in ground-state and lower-excitation baryons rather then in color-degenerate objects.

However, at this chemical potential the distance between quarks of neighboring baryons can be arbitrarily small in configuration space $\left( \sim 1/N_c \right)$, leading to the apparently paradoxical situation of quarks close enough to interact perturbatively (due to asymptotic freedom, with the scale given by configuration space inter-quark separation) in a confined medium.
\cite{quarkyonic} proposed to solve this conundrum by postulating that in the new phase the quarks below the Fermi surface act as free objects but the Fermi surface excitations are confined.  While the new phase is confined, the entropy density and pressure feels the quark degrees of freedom and $\sim N_c$, as the energy density.

The ``naive'' picture of this matter is that of overlapping nuclei where quarks can be freely exchanged by long-wavelength interactions, which also ensure color-neutrality at the scale of the nucleon size.   This would mean that the ``percolation''
picture of confinement \cite{perc1,perc2}, is wrong at high chemical potential:  A new state of matter exists where the ``percolation length'' order parameter for each quark diverges, yet the more conventional order parameters of confinement such as the Polyakov loop \cite{pol1,pol2} remain close to zero.
This new state of matter, called quarkyonic in \cite{quarkyonic}, should also be realized in our $N_c=3$ world and reachable in heavy ion collisions \cite{quarkyonicfit}  since large $N_c$ is at least qualitatively true in our world.

A great deal of investigation has gone on to see weather quarkyonic matter appears in any effective theory of QCD.   While a phase transition does seem to exist which has {\em some} of the characteristics described above \cite{sasaki1}, it is not clear weather the most interesting properties ($P \sim N_c$ and chiral symmetry restoration in the confined medium) are physically realized, as we do not have a model realistic enough but still computable.  Other approaches have found no evidence for any such transition \cite{baympaper,philipsenpaper},or have claimed the ``quarkyonic'' phase to have different properties for those claimed in \cite{quarkyonic} (eg \cite{satzquarkplasma} conjectures a chirally broken but deconfined constituent quark plasma).

As discussed in the introduction, the main difficulty of theoretical investigation in this regime is that there is no reliable approximation technique which is capable of distinguishing between models.
  The results obtained with these models, however, are highly dependent on the assumptions made in them, assumptions which can {\em not} be rigorously shown to derive uniquely from QCD.   In case of the critical point \cite{stephanov}, different models were shown to give very different answers.  Additionally, none of these models contain features unique to non-perturbative QCD, such as exact quark confinement.  As a consequence, the crucial aspect of the quarkyonic hypothesis, scaling of entropy density with $N_c$ in the quarkyonic phase, can not be adequately tested with models such as pNJL \cite{sasaki1}.

A possible way out are techniques deriving from Gauge-string duality \cite{maldacena}.  While no string theory with a dual looking like QCD is known, several models were developed which share with QCD some of its more notable non-perturbative characteristics,such as confinement and chiral symmetry breaking \cite{sugimoto}.    These models can be used to extrapolate to regions inaccessible to pQCD and the lattice, while retaining qualitative aspects of non-perturbative QCD such as its strongly coupled nature and dynamical confinement. 

 A finite chemical potential study \cite{lippert,rozali} within the Sakai-Sugimoto model \cite{sugimoto} has shown that the basic structure of the phase diagram is the same as Fig. \ref{phasediag}, and, just as in \cite{quarkyonic}, a new phase  II emerges, with the transition line at $\mu_q \sim \order{1}  N_c^0 \lqcd$ and the nuclear density as the order parameter, just like in \cite{quarkyonic}.
There are, however, profound differences:  \cite{lippert} finds that both phases I and II are confining and chiral-broken.    No evidence exists that the scaling of the pressure changes between I and II.   In fact, the only difference
between I and II seems to be a discontinuity in the Baryonic density.  The authors of \cite{lippert} interpret phase II as the well studied nuclear gas liquid phase transition \cite{liq0,liq1,liq3,liq5}, rather than as a new undiscovered phase.   If this interpretation is correct, than searching for the quarkyonic phase and/or the triple point separating I,II,III at upcoming low energy experiments \cite{low1,low2,low3,low4} would be fruitless, as in our $N_c=3$ world the liquid-gas phase has been extensively studied theoretically  and pinpointed experimentally, and its transition line is understood to lie well below $T_c$, so that no triple point exists.

These ambiguities reflect the persistent difficulties the large $N_c$ expansion has had in describing baryonic matter.
From the seminal work of \cite{witten}, it was understood that the baryon in the large $N_c$ limit is a semiclassical non-perturbative state, analogous to a skyrmion,where  $1/N_c$ then plays the role of a non-perturbative ``coupling constant''.
Nuclear matter,in this picture, becomes a ``skyrme crystal'' of tightly bound solitons \cite{crystal1,crystal2}.
The problem with this and subsequent works is that the resulting binding energy for nuclear matter is $\sim \order{N_c \lqcd} \sim \order m_{baryon}$.  This misses the realistic binding energy of nuclear matter by two orders of magnitude, making this picture of nuclear matter (a liquid) not even qualitatively correct.

This fact naively puts any extrapolations of nuclear matter based on large $N_c$ arguments under suspicion.   Simply saying such arguments are incorrect, however, is deeply unsatisfying:  $N_c=3$ {\em is large} in the sense that $N_c^2 \gg N_c$, so one {\em expects} the large $N_c$ picture to be qualitatively correct with $\order{30\%}$ quantitative corrections.
Indeed, at zero chemical potential this seems to work remarkably well \cite{thooft,witten}, and much better then expected when precision (lattice) calculations are performed \cite{lat1,lat2}.   It is therefore simply not good enough to say that ``large $N_c$ does not work'' unless a convincing physical reason is offered as for why.   This work aims to conjecture such a reason.

\section{The ``large $N_c$ limit'' in a densely packed system}

We conjecture that the large $N_c$ description of nuclear matter is flawed because $N_c=3$, while roughly $\gg 1$, is smaller than the other dimensionless scale relevant at high density:  $N_N$, the number of neighbors a nucleon has in a tightly packed nuclear material.   The more neighbors, the more Pauli blocking of valence quarks must be important, and the more the presence of neighbors will disturb the configuration space part of the quark wavefunction inside the nucleons.

Since,due to the uncertainty principle, any such disturbance of the nuclear wavefunction adds an energy of the order of the confinement scale $\sim \lqcd$, the nuclear repulsive core will be larger than the inverse of the nuclear separation up to the deconfinement temperature.  If the number of colors is larger than $N_N$, this problem will not exist since it will be possible to arrange the color part of the wavefunction so the nearest quarks of neighboring baryons will be of different colors. In this limit baryons can be tightly packed (interbaryonic separation $\sim \lqcd$) without the configuration space part of the baryonic wavefunction being disturbed.   

Thus, the limit in which exciting baryonic resonances is ``cheap'' ($\Delta E\sim 1/N_c$), first suggested in \cite{witten} and used in \cite{quarkyonic} to argue why entropy density scales as energy density $\sim N_c$ in dense baryonic matter, is only valid when $N_c \gg N_N$.
$N_N$, of course, is a function not of $N_c$, but the (fixed) number of dimensions $d$ and ``packing scheme'', $N_N \sim  k(d) N_c^0$. The exact form of the ``kissing number'' function $k(d)$ in arbitrary dimensions is unknown \cite{kissing}, but seems to be approximated by $k(d) \sim 2^{\alpha d}$, with a transcendental $\alpha\simeq 0.22$ .      $k(1,2,3)$ is, respectively, 2,6 and 12.

If our conjecture is correct, it becomes plausible that, while the ``quarkyonic phase'' is reasonable in an $N_c \gg k(3)$ system, it is not so in our world.
 If $N_c \ll k(d)$, as in our world, the Pauli exclusion principle keeps the
nuclear excluded volume at a value significantly larger than $\lqcd^{−3}$. In this case, confinement suppresses the exchange
of colored degrees of freedom between the nuclei, so the entropy carried by inter-nuclear forces $\sim N_c^0$. It would also mean that the percolating phase transition studied in \cite{perc1,perc2} coincides with deconfinement.

In a $N_c \gg k(d)$ world, however, nuclei touch each other, and colored degrees of freedom can freely percolate
between them. The entropy carried by these percolating degrees of freedom $\sim N_c$, and in the large color limit ends up
overwhelming the total entropy of the system, in much the same way that the electron gas carries most of the entropy
of a metal (Note that the equilibrium entropy of colored objects $\sim N_c$ even if interaction cross-section between these objects is $N_c$-suppressed. The {\em timescale of equilibration} gets longer, but the equilibrium entropy stays the same).   In this limit, the percolation transition \cite{perc1,perc2} does not represent deconfinement but the quarkyonic transition, and the two, in chemical potential, are separated by  $\Delta \mu \sim N_c \lqcd$.

\begin{figure}[h]
\includegraphics[scale=0.4]{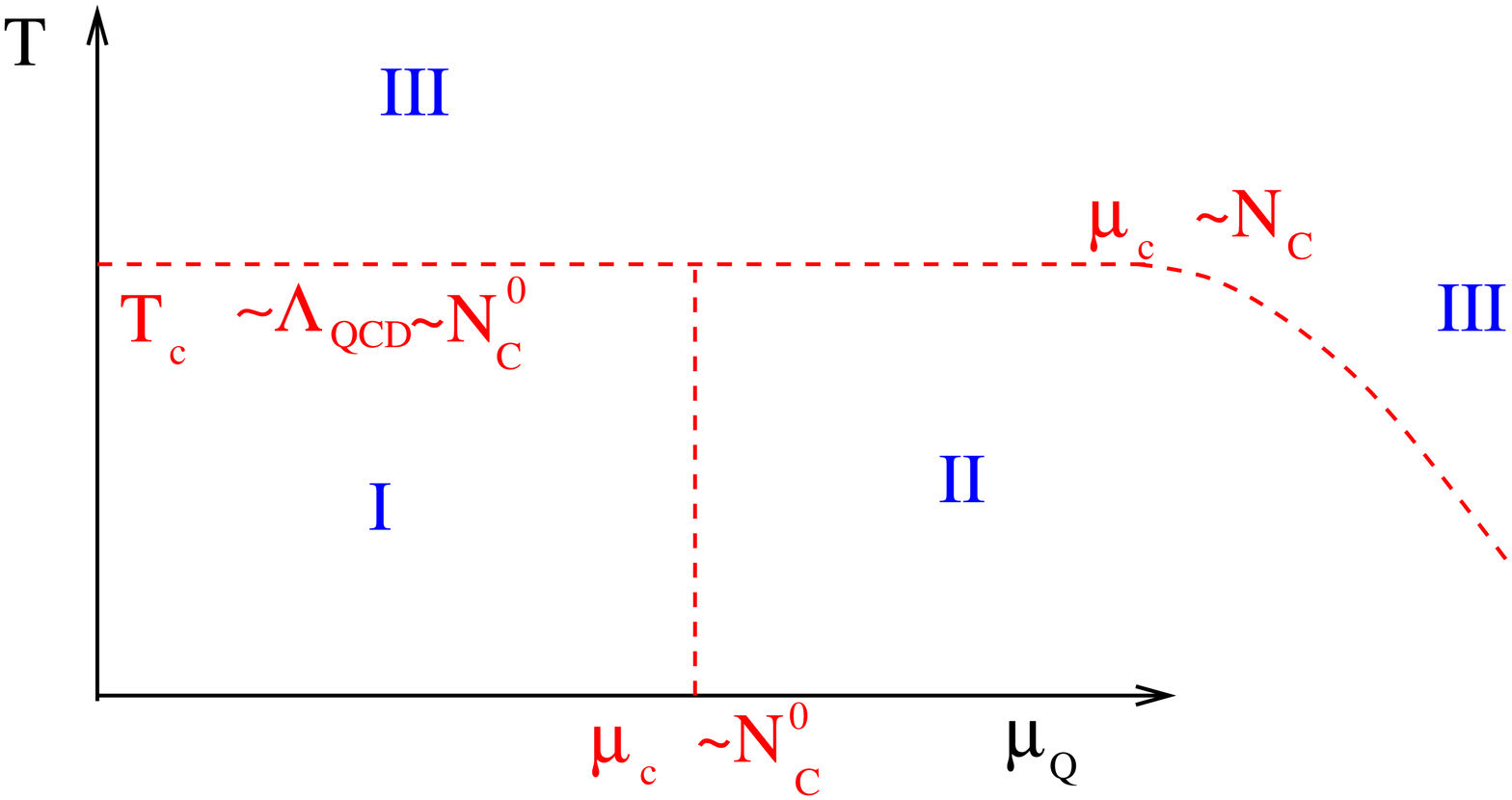}
\caption{(color online)The phase diagram for large $N_c$. 
See text for a description of the phases I,II,III in various models. 
\label{phasediag} }
\end{figure}

While this conjecture is reasonable, testing it in a systematic manner is an upcoming research project that will take some time to complete \cite{lottini}.  As a first step \cite{largenc}, we can show that, when the parameters of the Van Der Waals gas model are varied with $N_c$ according to the prescription given here, the phase diagram interpolates between the usually accepted nuclear matter phase diagram and one which is very similar to Fig. \ref{phasediag}.

In the large $N_c$ limit, the only $N_c$-invariant scale of the theory is $\lqcd$, the scale at which the 't Hooft coupling constant becomes $\lambda \sim \order{1}$.   While a precise value of this scale depends on the scheme used to calculate it, its roughly $\lqcd \sim N_c^0 \simeq 200-300$ MeV \cite{pdg}.
It is therefore natural to expect that any physical quantity is $\sim f(N_c) \lqcd^d$, a dimensionless function of $N_c$ times a power of $\lqcd$ set by the dimensionality $d$ of the quantity.
Henceforward we shall adopt this assumption, and, for brevity, set $\lqcd$ to unity in the equations.    The reader should multiply any dimensionful quantity in the equations by the appropriate power of $\lqcd$ (For example, the Baryon mass is $\sim N_c \lqcd$ in the text, and $\sim N_c$ in the equations).
In this notation, the Van Der Waals parameters $a$, $b$ and the curvature correction become dimensionless $\alpha$,$\beta$,$\gamma$ times the appropriate power of $\lqcd$ (3 for $\alpha$,2 for $\beta$,4 for $\gamma$), and the VdW equation \cite{chemref2} becomes
\begin{equation}
\left( \rho^{-1} - \alpha \right) \left(  P + \beta \rho^2 - \gamma  \rho^3 \right) = T
\label{vdw}
\end{equation}
The considerations in our previous paragraph lead us to assume that
\begin{equation}
\label{eqalpha}
\alpha \sim \order{ \frac{N_N}{N_c}} + 1 \sim \order{ \frac{k(d)}{N_c}}+1  \sim \left.  \order{ \frac{10}{N_c}}+1  \right|_{d=3}
\end{equation}
The coefficients $\beta,\gamma$ should, according to \cite{witten,crystal1,crystal2} go as $N_c$.   Recent work \cite{larrynucleon}, however, has cast doubt on this assumption and proposed they go as $\sim N_c^0$ or $\sim \ln N_c$.

The chemical potential can be obtained \cite{chemref2} by the textbook thermodynamic relation
$\rho = \left(dP/d\mu  \right)_T$.   Inverting, and writing in terms of $\mu_q=\mu_B/N_c$ we have
\begin{equation}
\label{mubexact}
\mu_q = 1 +\frac{1}{N_c} \left[ \int^\rho_0 f(\rho',T) d\rho' + F(T)  \right]
\end{equation}
where the first term is the nucleon mass and 
\begin{equation}
\label{fdef}
f(\rho,T)= \left( \frac{ dP}{d \rho} \right)_{T} \frac{1}{\rho} = \frac{T}{ \rho  (1-\alpha \rho )^2}+2\beta
\end{equation}
$\rho$ and $P$ are the density at the phase transition, which could be liquid  $\rho_l$ or gas $\rho_g$ (if the calculation is performed correctly the same chemical potential should come out).
$\rho_{l,g}$ are in turn the solutions to the equation \ref{vdw} in the region where this equation has two solutions.
Obtaining all such solutions is trivial at the mathematical level through algebraically cumbersome.  The reader can get the detailed results in \cite{largenc}.

\begin{figure}[h]
    \centering
        \includegraphics[width=0.4\textwidth]{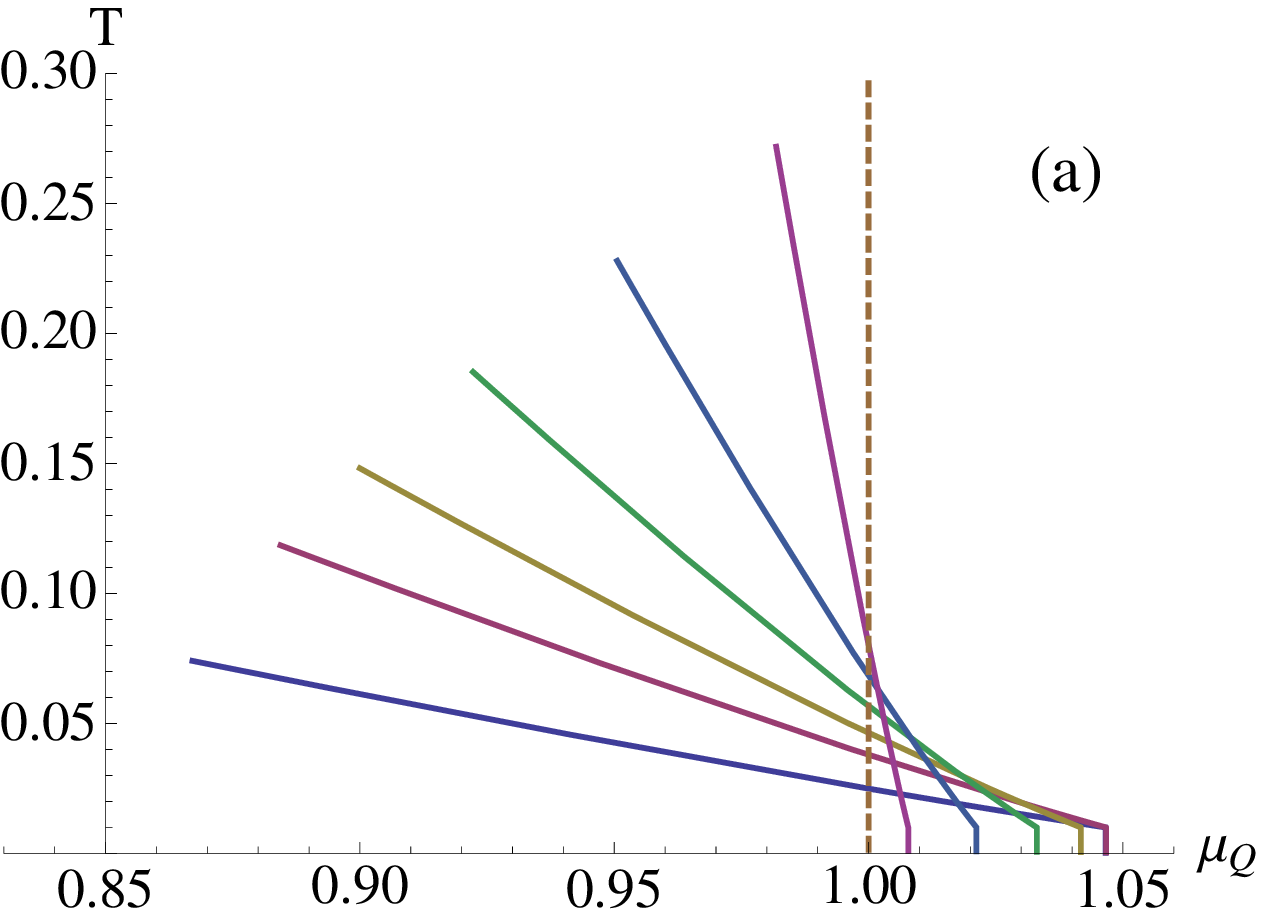}
\includegraphics[scale=0.6]{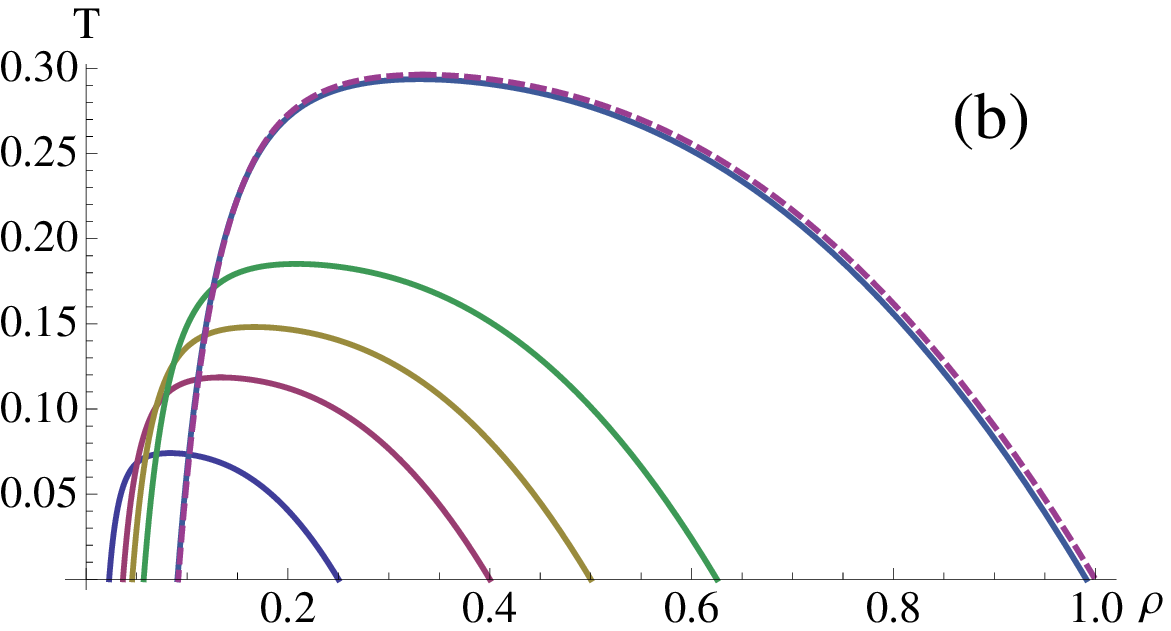}
       \includegraphics[width=0.4\textwidth]{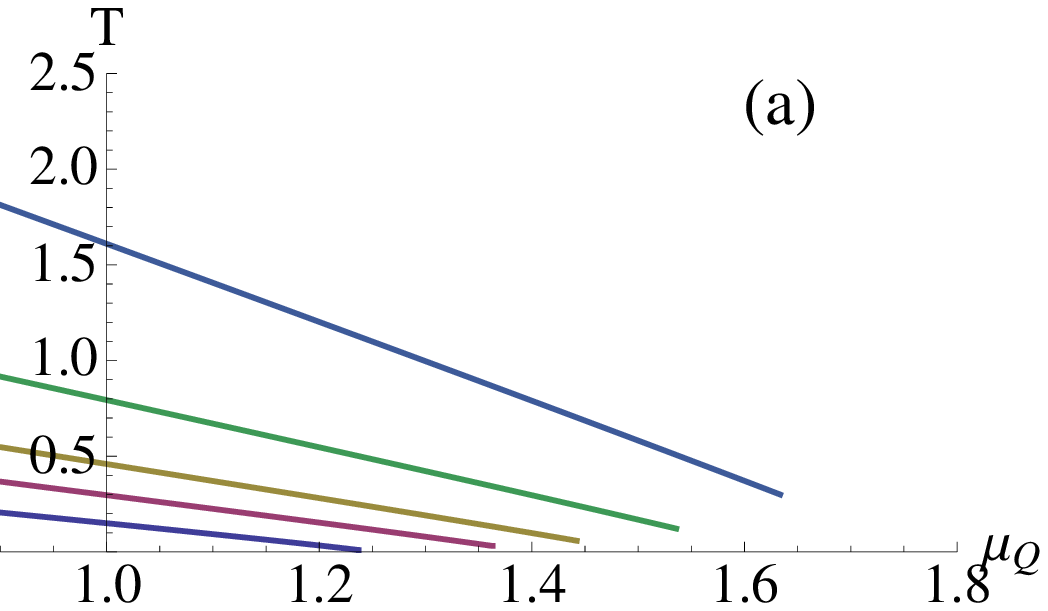}
\includegraphics[scale=0.6]{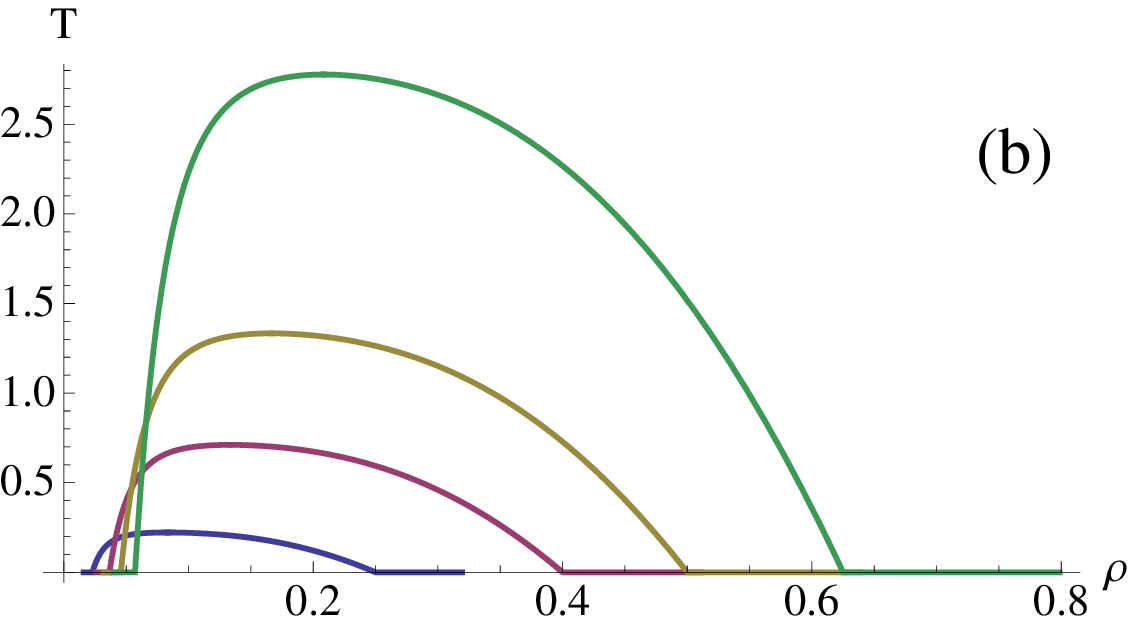}
\caption{(color online) The phase diagram in $T-\mu$ space (left panel) and $T-\rho$ space (right panel) as a function of $N_c$ ($N_c=3,5,8,10,30,100$, with increasing color corresponding to a line with higher $T,\mu,\rho$)
Top panels assume nuclear interactions $\sim N_c^0$ or $\sim \ln N_c$, bottom panels as $\sim N_c$ }
\label{phasemu}
\end{figure}

The result is shown in Fig. \ref{phasemu}.  As can be seen, for $N_c \ll N_N$, the phase diagram looks qualitatively similar to the liquid-gas phase transition in our world, \cite{liq0,liq1,liq3,liq5}, with the phase transition line close to horizontal and $T_c \ll \lqcd$.  In the opposite limit, $N_c \gg N_N$, $T_c \order \lqcd$ and the phase diagram becomes nearly vertical provided $\beta,\gamma \sim N_c^0$ or $\sim \ln N_c$.  If $\beta,\gamma \sim N_c$, the curvature of the phase transition does not go to zero and $T_c \sim N_c \lqcd$, which makes it go above the deconfinement transition.
This signals that, if $\beta,\gamma \sim N_c$ the Van Der Waals approach breaks down at large $N_c$, a result natural if the description in \cite{crystal1,crystal2} is correct in this limit.  In this case too, through, the low $N_c$ limit is parametrically close to the nuclear liquid-gas world ($T_c \ll \lqcd$ and $\rho_g,\rho_l \ll \lqcd^3$), and as $N_c$ increases $\rho_{g,l}$ goes to its critical value $\lqcd^3$ where nuclei overlap.  While in this limit the curvature of the phase transition does {\em not} go to zero, this {\em might} be an artifact of a badly broken approximation.

In conclusion, we reported on the ambiguities in our understanding of nuclear matter at moderate temperature and chemical potential ($\sim \lqcd$), particularly in regards to the extrapolations at large $N_c$.
We have argued that some of this ambiguity comes from the large differences between the expectation of large $N_c$ QCD and the experimental nuclear ground state.  We have conjectured that this is due to the fact that the true ``large $N_c$ limit'' for dense matter comes when $N_c \gg \order{10}$, the number of neighbors in a closely packed system, and shown that when this conjecture is implemented in the Van Der Waals nuclear gas, limits looking like the real world and the large $N_c$ limit seem to emerge.  To explore this conjecture in rigorous and systematic way requires further work.
\begin{acknowledgments}
IM acknowledges support provided by the DFG grant 436RUS 113/711/0-2 (Germany) and grant NS-7235.2010.2 (Russia). G.T. acknowledges the financial
support received from the Helmholtz International Center for FAIR within the framework of the LOEWE program (Landesoffensive zur Entwicklung
Wissenschaftlich-\"Okonomischer Exzellenz) launched by the State of Hesse.  G.~T. thanks M.~Gyulassy and Columbia University for the hospitality
provided when part of this work was done. The authors thank Jorge Noronha, Larry McLerran,Rob Pisarski,Stefano Lottini and Krystof Redlich for discussions
\end{acknowledgments}

\newpage

\end{document}